\documentclass[twocolumn,showpacs,preprintnumbers,amsmath,amssymb]{revtex4}

\usepackage{bm}
\usepackage{epsfig}
\usepackage{fancyheadings}
\usepackage{fullpage}
\usepackage{amsmath}
\usepackage{amsbsy}
\usepackage{amssymb}
\usepackage{amsthm}
\usepackage{amscd}
\usepackage{amsfonts}
\usepackage{supertabular}
\usepackage{graphics}
\usepackage{graphicx}
\usepackage{verbatim}
\usepackage{subfigure}
\usepackage{epsfig}
\usepackage{xspace}
\usepackage{euscript}
\usepackage{alltt}
\usepackage{boxedminipage}
\usepackage{float}
\usepackage{times}

\newcommand{\mbs}[1]{\boldsymbol{#1}}
\newcommand{\mbb}[1]{\mathbb{#1}}

  \def\fb{{\mbs{f}}}

 \def\bq{{\mbs{q}}}

\def\dt{{\triangle t}}

\begin{document}

\title{Statistical mechanics of dissipative systems}

\author{M.~Koslowski and M.~Ortiz\footnote{Author to whom correspondence should be addressed}}\email{ortiz@aero.caltech.edu}
\affiliation{Engineering \& Applied Science Division \\
        California Institute of Technology \\
        Pasadena, CA 91125, USA\\
(Submitted to Physical Review Letters)}


\begin{abstract}

We propose a generalization of classical statistical mechanics
which describes the behavior of dissipative systems placed in
contact with a heat bath. In contrast to conventional statistical
mechanics, which assigns probabilities to the \emph{states} of the
system, the generalized theory assigns probabilities to the
\emph{trajectories} of the system. The conditional probability of
pairs of states at two different times is given by a path integral.
We present two simple analytically-tractable examples which
illustrate the predicted effect of temperature on the mean
trajectories, hysteresis and drift of the system.

\end{abstract}



\maketitle

The fundamental question addressed in the present communication
is: \emph{How does a dissipative system behave when it is placed
in contact with a heat bath?} It should be carefully noted that
the systems under consideration are dissipative {\it ab initio},
e.~g., as a result of internal friction, viscosity, or some other
dissipative mechanism. In particular, the systems are
irreversible, path dependent, and exhibit hysteresis. Therefore,
the objective of the theory is \emph{not} to understand how
dissipation arises from the coarse-graining of a conservative
system or in connection with first-order phase transitions (see
\cite{Jiles:1984, Ortin:1991, sethna:1993, kinderlehrer:1997} for
notable examples of that line of inquiry). Instead, the focus is
in understanding how a heat bath influences the behavior of a
dissipative system; what are the statistical properties of its
trajectories; and how the effective behavior depends on
temperature.

For definiteness, we consider systems whose state is defined by an
$N$-dimensional array $\bq$ of generalized coordinates. The
energetics of the system is described by an energy function
$E(\bq, t)$. The explicit time dependence of $E$ may arise, e.~g.,
as a result of the application to the system of a time-dependent
external field. In addition, the system is assumed to possess
\emph{viscosity}, and, thus, the equilibrium equations are of the
form:
\begin{equation}\label{eq:Equilibrium}
\fb \equiv \frac{\partial E}{\partial \bq}(\bq, t) + \fb^{\rm
vis}(\dot\bq) = 0
\end{equation}
where $\fb^{\rm vis}$ are the viscous forces. These equilibrium
equations define a set of ordinary differential equations which,
given appropriate conditions at, e.~g., $t=0$, can be solved for
the trajectory $\bq(t)$, $t \geq 0$.

If the system is conservative, i.~e., if $\fb^{\rm vis} = {\bf
0}$, the instantaneous state $\bq(t)$ of the system at time $t$
follows directly from energy minimization, i.~e., from the
problem:
\begin{equation}\label{eq:MinE}
E\big(\bq(t), t\big) = \min_{\bq' \in {\mbb{R}^N} } E(\bq', t)
\end{equation}
In order to extend this variatonal framework to dissipative
systems, we resort to time discretization, leading to a
\emph{sequence} of minimization problems of the form
(\ref{eq:MinE}) \cite{OrtizRepetto1999, RadovitzkyOrtiz1999,
OrtizStainier1999}. Thus, we consider a \emph{time-discretized}
incremental process consisting of a sequence of states ${\bf
q}_{n-1}$ at times $t_0=0$, $\dots$, $t_{n-1}$, $t_n = t_{n-1} +
\dt$, $\dots$. We additionally assume that the viscous forces
$\fb^{\rm vis}$ derive from a kinetic potential $\phi(\dot\bq)$
through the relation:
\begin{equation}
\fb^{\rm vis} = \frac{\partial\phi}{\partial \dot{\bf q}}(\dot\bq)
\end{equation}
and introduce the incremental work function:
\begin{equation}\label{W}
\begin{split}
    & W(\bq_n, \bq_{n-1}) = E(\bq_n) - E({\bf
    q}_{n-1}) \\
    & + \min_{\{paths\}} \int^{t_n}_{t_{n-1}} \phi\big( \dot{\bf
    q}(t) \big) \, dt
\end{split}
\end{equation}
where the minimum is taken over all paths $\bq(t)$ such that
$\bq(t_{n-1}) = \bq_{n-1}$ and $\bq(t_n) = \bq_n$. The
fundamental property of the incremental work function $W({\bf
q}_n, \bq_{n-1})$ is that it acts as a potential for the forces
$\fb_n$ at time $t_n$, i.~e.,
\begin{equation}
\fb_n= \frac{\partial W}{\partial\bq_n}(\bq_n, {\bf q}_{n-1})
\end{equation}
In order to verify this property, we may consider a small
perturbation $\bq_n \to \bq_n + \delta \bq_n$, leading to a
corresponding perturbation of the path $\bq(t) \to \bq(t) + \delta
\bq(t)$, with $\delta \bq(t_{n-1}) = \delta \bq_{n-1} = {\bf 0}$.
Then, in follows that
\begin{equation}
\begin{split}
    & \delta W(\bq_n, \bq_{n-1}) = \frac{\partial
    E}{\partial \bq}(\bq_n) \cdot \delta \bq_n +
    \fb^{\rm vis}_n \cdot \delta \bq_n \\
    & + \int^{t_n}_{t_{n-1}} \fb^{\rm vis}\big( \dot\bq(t) \big)
    \cdot \delta \dot\bq(t) \, dt
\end{split}
\end{equation}
But the integral on the right hand side of this equation is to be
evaluated along the minimizing path, and hence it vanishes
identically. This gives the identity
\begin{equation}
\delta W(\bq_n, \bq_{n-1}) = \left\{ \frac{\partial E}{\partial
\bq}(\bq_n) + \fb^{\rm vis}_n \right\} \cdot \delta {\bf q}_n
\end{equation}
Since the variation $\delta \bq_n$ is arbitrary, it follows that
\begin{equation}
\frac{\partial W}{\partial \bq_n}(\bq_n, \bq_{n-1}) =
\frac{\partial E}{\partial \bq}(\bq_n)
 + \fb^{\rm vis}_n = \fb_n
\end{equation}
as stated. From this property it follows that the equilibrium
equation $\fb_n = {\bf 0}$ is the Euler-Lagrange equation
corresponding to the minimum principle:
\begin{equation}\label{MinW}
\min_{\bq_n} W(\bq_n, \bq_{n-1})
\end{equation}
This behavior is indistinguishable from that a conservative system
with `energy' $W(\bq_n , \bq_{n-1})$. However, it should be
carefully noted that $W$ is determined by both the
\emph{energetics} and the \emph{kinetics} of the system.
Consequently, $W$ depends on the initial conditions $\bq_{n-1}$
for the time step and, therefore, varies from step to step, which
in turn allows for path dependency and hysteresis, as required. A
simple situation arises when the kinetic potential
$\phi(\dot{\bq})$ is convex and coercive, i.~e., it grows as
$|\dot{\bq}|^p$, for some $p \in (1, \infty)$, for large
$|\dot{\bq}|$. Then, the minimizing path is unique and consists of
a straight segment joining $\bq_{n-1}$ and $\bq_n$. Under these
conditions, the corresponding incremental work function is
\begin{equation}\label{W2}
W(\bq_n, \bq_{n-1}) = E(\bq_n) - E(\bq_{n-1}) + \triangle t \,
\phi\left( \frac{\triangle\bq}{\triangle t} \right)
\end{equation}
where we write $\triangle \bq = \bq_n - \bq_{n-1}$.

Now, imagine placing the system just described in contact with a
heat bath at absolute temperature $T$. The objective is to predict
the ensuing behavior of the system. For conservative systems, such
behavior is predicted by standard Gibbsian statistical mechanics,
to wit: the probability of finding a state $\bq$ at time $t$ obeys
Boltzmann's distribution, i.~e., is proportional to $\exp\{-\beta
E(\bq, t)\}$, where $\beta = 1/k_B T$ and $k_B$ is Boltzmann's
constant. In particular, the states of the system at two different
times are uncorrelated. We proceed to show that the analogy
between the minimum principles (\ref{eq:MinE}) and (\ref{MinW})
provides a basis for generalizing Gibbs's prescription to
dissipative systems. However, in contrast to conventional
statistical mechanics, which assigns probabilities to the
\emph{states} of the system, the extended theory assigns
probabilities to the \emph{trajectories} of the system.

We begin by assuming that the incremental processes are Markovian,
i.~e., $\bq_n$ is correlated to $\bq_{n-1}$ but not to earlier
states. Let $p(\bq_n, \bq_{n-1})$ be the joint probability of
$\bq_n$ and $\bq_{n-1}$, and let
\begin{subequations}
\begin{align}
    & p(\bq_{n-1}) = \int p(\bq_n,\bq_{n-1})
    \, d^Nq_n \\
    & p(\bq_n) = \int p(\bq_n,\bq_{n-1})
    \, d^Nq_{n-1}
\end{align}
\end{subequations}
be the corresponding probabilities of $\bq_n$ and ${\bf
q}_{n-1}$. Then the conditional probability of $\bq_n$ given
$\bq_{n-1}$ is
\begin{equation}
p(\bq_n|\bq_{n-1} ) = \frac{p(\bq_n,\bq_{n-1})}{p({\bf q}_{n-1})}
\end{equation}
This probability may also be interpreted as the transition
probability from state $\bq_{n-1}$ to the new state ${\bf q}_n$.
\emph{We postulate that the transition probability
$p(\bq_n|\bq_{n-1})$ is Gibbsian}, i.~e.,
\begin{equation}\label{eq:Trans}
p(\bq_n|\bq_{n-1} ) = \frac{1}{Z(\bq_{n-1})}e^{-\beta W({\bf
q}_n,\bq_{n-1})}
\end{equation}
where
\begin{equation}\label{eq:Z}
Z(\bq_{n-1})= \int e^{-\beta W(\bq_n,\bq_{n-1})} \, d^Nq_n
\end{equation}
is an incremental partition function. It therefore follows that:
\begin{equation}
p(\bq_n) = \int \frac{1}{Z(\bq_{n-1})}e^{-\beta W({\bf
q}_n,\bq_{n-1})} p(\bq_{n-1}) \, d^Nq_{n-1}
\end{equation}
Iterating this relation we obtain
\begin{equation}\label{eq:pq}
\begin{split}
    & p(\bq_n) = \\
    & \int \dots \int \left \{ \prod_{i=0}^{n-1}\frac{e^{-\beta
    W(\bq_{i+1},\bq_{i}) }}{Z(\bq_i)} \right \} \,
    p(\bq_0) \, d^Nq_{n-1} \dots d^Nq_{0}
\end{split}
\end{equation}
In the limit of $\dt \to 0$ and $n \to \infty$ at fixed $t = n
\dt$, (\ref{eq:pq}) defines a path integral of the form (see,
e.~g., \cite{Wiegel:1986}):
\begin{equation}\label{eq:path}
p(\bq_t | \bq_0) = \int \dots \int e^{-\beta \EuScript W[\bq]}
\EuScript D q
\end{equation}
for some system-dependent work functional $\EuScript W[\bq]$. This path
integral gives the conditional probability $p(\bq_t | \bq_0)$ of
finding the system in state $\bq_t$ at time $t \geq 0$ given that
the system is in state $\bq_0$ at time $t=0$. The integrand of
(\ref{eq:path}) may be interpreted as relating to the probability
of individual trajectories $q(\tau)$. More precisely, given two
trajectories $\bq_1(\tau)$ and $\bq_2(\tau)$, their relative
probabilities are:
\begin{equation}
\frac{p[\bq_1]}{p[\bq_2]} = \frac{e^{-\beta W[\bq_1]}}{e^{-\beta
W[\bq_2]}}
\end{equation}
It is noteworthy that, in contrast to the conservative case, the
presence of kinetics renders states of the system at different
types \emph{correlated}. However, this correlation may be expected
to decay with elapsed time, conferring the system a \emph{fading
memory} property.

In the remainder of this letter we present two simple
analytically-tractable examples which provide a first illustration
of the theory, namely: a linear spring-dashpot system; and dry
friction. In addition to their value as illustrative examples,
these simple models should also useful as a basis for constructing
\emph{mean-field} approximations to more complex systems. The
treatment of general systems requires the use of numerical schemes
such as the Path-Integral Monte Carlo (PIMC) method \cite{Binder:1992}.

\begin{figure}
\epsfig{file=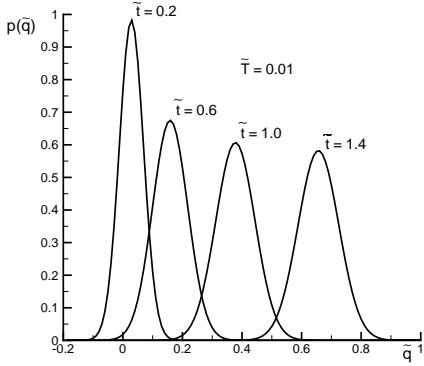,height=0.25\textheight}
\caption{Evolution of the probability density for a spring and
dashpot system subjected to time-proportional loading.}
\label{fig:vis-lin-10}
\end{figure}

The linear spring-dashpot system is characterized by an energy
function and kinetic potential of the form:
\begin{subequations}
\begin{align}
    & E(q, t) = \frac{\kappa}{2} q^2 - f(t) q \\
    & \phi(\dot{q}) = \frac{\eta}{2} \dot{q}^2
\end{align}
\end{subequations}
where $q \in \mbb{R}$, $\kappa$ is the spring constant, $\eta$ is
the dashpot viscosity, and $f(t)$ is a time-dependent applied
force. For this system, the equilibrium equations
(\ref{eq:Equilibrium}) reduce to:
\begin{equation}\label{eq:Equilibrium1D}
\kappa q - f + \eta \dot{q} = 0
\end{equation}
The corresponding incremental work function for this system is:
\begin{equation}
\begin{split}
    & W(q_n, q_{n-1}) = \frac{\kappa}{2} (q_n^2 - q_{n-1}^2) \\
    & - f_n q_n + f_{n-1} q_{n-1} + \frac{\eta}{2\dt} (q_n - q_{n-1})^2
\end{split}
\end{equation}
and the evaluation of the partition function entails a simple
Gaussian integral. Inserting the result in (\ref{eq:pq}) yields
\begin{equation}\label{eq:pq1D}
\begin{split}
    & p(q_n) =
    \int \dots \int \left( \frac{2 \pi}{\beta (\kappa + \eta/\dt)}
    \right)^{-n/2} \\
    & \exp\left\{ -\beta \sum_{i=0}^{n-1}
    \frac{\big( \dt (f_{i+1} - \kappa q_{i+1}) -
    \eta (q_{i+1} - q_i) \big)^2}
    {2 \dt (\eta + \kappa \dt) } \right\} \\
    & p(q_0) \, dq_{n-1} \dots dq_{0}
\end{split}
\end{equation}
In the limit of $\dt \to 0$ and $n \to \infty$, with $t = n\dt$
fixed, the sum in the exponential of this formula converges to the
Riemann integral
\begin{equation}\label{eq:WFunc1D}
    \EuScript W[q] =
    \int_0^t \frac{1}{2\eta} \big( \kappa q(\tau) - f(\tau) +
    \eta \dot{q}(\tau) \big)^2 \, dt
\end{equation}
In the same limit, (\ref{eq:pq1D}) may be written as the path
integral
\begin{equation}\label{eq:Dq1D}
    \EuScript D q = \lim_{n \to \infty}
    \left (\frac{2 \pi}{\beta\big(\kappa +
    \eta/(t/n)\big)}\right )^{-n/2} \,dq_{n-1} \dots dq_{0}
\end{equation}
which is of the anticipated form (\ref{eq:path}). The structure of
(\ref{eq:path}) is revealing. Thus, it is observed that the work
functional $\EuScript W[q]$ integrates in time the square of the
equilibrium equation (\ref{eq:Equilibrium1D}), reduced to units of
power by means of the factor $1/\eta$. Evidently, the path which
contributes the most to the path integral is the \emph{critical
path}, i.~e., the path $q_c$ which minimizes the work functional
(\ref{eq:WFunc1D}). Indeed, in the limit of zero viscosity or zero
temperature, the \emph{only} path which contributes to the path
integral is the critical path, which in that limit coincides with
the deterministic  trajectory, i.~e., with the solution of
(\ref{eq:Equilibrium1D}). For finite viscosity and finite
temperature, however, all paths contribute to the path integral to
varying degrees, the contributions becoming increasingly weaker as
the trajectories depart from the critical path. The stationarity
of $ \EuScript W[q]$ yields the Euler-Lagrange equations for the critical
path
\begin{equation}\label{eq:Equilibrium2}
\ddot{q} - \omega^2 q = \frac{\dot{f}}{\eta} - \omega^2
\frac{f}{\kappa}
\end{equation}
where $\omega = \kappa/\eta$ and the solution $q_c$ is subject to
the boundary conditions $q_c(0) = q_0$, $q_c(t) = q_t$. It is
interesting to note that this equation is of second order in time,
whereas the original equilibrium equation (\ref{eq:Equilibrium1D})
is of first order. This increase in order makes it possible to
enforce boundary conditions at times $0$ and $t$, in contrast to
the original equation (\ref{eq:Equilibrium1D}) which only allows
for initial conditions. Furthermore, we note that
(\ref{eq:Equilibrium2}) is obtained by \emph{squaring}
(\ref{eq:Equilibrium1D}), and thus the solutions of the latter are
subsumed within the former. However, by virtue of its higher order
the Euler-Lagrange equation (\ref{eq:Equilibrium2}) has solutions
which do not satisfy (\ref{eq:Equilibrium1D}). It thus follows
that the critical path does not coincide with the deterministic
trajectory in general. Conveniently, the path integral
(\ref{eq:path}) can be evaluated exactly in closed form (e.~g.,
\cite{Wiegel:1986} and \cite{Feynman:1990}). The result is:
\begin{equation}
p(q_t) =  \sqrt{ \frac{\beta \kappa e^{\omega t}}{ 2 \pi
\sinh(\omega t) } } e^{-\beta ( \EuScript W[q_c] )}
\label{eq:p(q)}
\end{equation}
where the critical path $q_c(t) $ is the solution of
(\ref{eq:Equilibrium2}). Fig.~\ref{fig:vis-lin-10} shows the
evolution of the probability density for a linear loading history
$f(t) = A t$ in terms of the normalized variables $\tilde{q} = q /
( A \eta / \kappa^2 )$, $\tilde{t} = t / ( \eta/ \kappa )$, and
$\tilde{T} = T / ( A^2 \eta^2 / k_B \kappa^3 )$. The mean value
$\langle q(t) \rangle$ traces the critical path $q_c(t)$, which,
as noted earlier, differs from the classical path at finite
viscosity and temperature. An additional effect of viscosity,
which is clearly evident in these figures, is to cause the width
of the probability density to broaden in time, with the attendant
increase in the uncertainty of the state of the system.

\begin{figure}
\epsfig{file=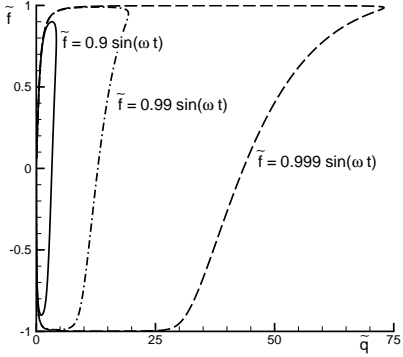,height=0.25\textheight}
\caption{Trajectories of a system sliding under the action of an
applied cyclic load $f(t) = A \, \sin(\omega t)$ against a
frictional resistance $s$.} \label{fig:dry}
\end{figure}

The work functional (\ref{eq:WFunc1D}) appearing in the path
integral representation (\ref{eq:path}) is characteristic of
systems whose kinetic potential $\phi(\dot{\bq})$ exhibits
quadratic behavior near the origin. However, there are cases of
interest which do not fall into that category. A case in point is
provided by \emph{dry friction}, which is characterized by a
kinetic potential which has a vertex at the origin. Consider, by
way of example, a one-dimensional system sliding against a
frictional resistance $s > 0$ under the action of an applied force
$f(t)$ such that $|f(t)| \leq s$. In this particular case
\begin{subequations}
\begin{align}
    & E(q, t) = - f(t) q \\
    & \phi(\dot{q}) = s |\dot{q}|
\end{align}
\end{subequations}
and the incremental work function (\ref{W}) specializes to
\begin{equation}
W(q_n , q_{n-1}) = -f_n q_n + f_{n-1} q_{n-1} + s \, | q_n -
q_{n-1} |
\end{equation}
As expected from the rate-independent nature of dry friction, $W$
is independent of $\triangle t$. The corresponding partition
function (\ref{eq:Z}), transition probability (\ref{eq:Trans}),
and probability density (\ref{eq:pq}) take the form:
\begin{equation}
Z(q_{n-1}) = \frac{ 2 s }{ \beta ( s^2 - f_n^2 ) } {\rm e}^{
\beta ( f_n - f_{n-1} ) \, q_{n-1} }
\end{equation}
\begin{equation}\label{eq:DF:Trans}
p( q_n | q_{n-1} ) =  \frac{\beta (s^2 - f_{n-1}^2)}{2 s} {\rm
e}^{ s \, | q_n - q_{n-1} | - f_n ( q_n - q_{n-1} ) }
\end{equation}
and
\begin{equation}
\begin{split}
    & p(q_n) =
    \int \dots \int \prod_{i=0}^{n-1} \frac{\beta (s^2 - f_i^2)}{2 s} \\
    & \exp\left\{ -\beta \sum_{i=0}^{n-1} \big( s \, | q_{i+1} - q_i | -
    f_{i+1} ( q_{i+1} - q_i ) \big) \right\} \\
    & p(q_0) \, dq_{n-1} \dots dq_{0}
\end{split}
\end{equation}
respectively. An important feature of the transition probability
(\ref{eq:DF:Trans}) is that it depends solely on the difference
$q_n - q_{n-1}$. Therefore, the chain $q_0$, $q_1$, $\dots$, $q_n$
defines a random walk, and the probability $p(q_t | q_0)$
corresponding to the limit of $\dt \to 0$ and $n \to \infty$ at
constant $t = n \dt$ follows by an application of the central
limit theorem. The result is:
\begin{equation}
p(q_t | q_0) = \frac{1}{\sqrt{2\pi} \sigma(t)} {\rm e}^{-\big( q_t
- \langle q \rangle(t) \big)^2/2\sigma^2(t)}
\end{equation}
where
\begin{equation}
\begin{split}
    \langle q \rangle(t) & = q_0 + \lim_{n\to\infty} \frac{1}{n}
    \sum_{i=0}^{n-1} \frac{2 f_i}{\beta ( s^2 - f_i^2 )} \\
    & = q_0 + \frac{1}{t} \int_0^t \frac{2 f(\tau)}
    {\beta \big( s^2 - f^2(\tau)\big) } \, d\tau
\end{split}
\end{equation}
is the mean path traced by the system, and
\begin{equation}
\begin{split}
    \sigma^2(t) & = \lim_{n\to\infty} \frac{1}{n} \sum_{i=0}^{n-1}
    \frac{2 (s^2 + f_i^2)}{\beta^2 (s - f_i)^2(s + f_i)^2} \\
    & = \frac{1}{t} \int_0^t \frac{2 \big(s^2
    + f^2(\tau)\big)}{\beta^2 \big( s - f(\tau) \big)^2
    \big( s + f(\tau) \big)^2}
\end{split}
\end{equation}
measures the deviation from the mean path. It is interesting to
note that, in the limit of zero temperature, $\langle q \rangle(t)
= q_0$ for as long as $|f(t)| < s$, with the system sliding off to
$\pm\infty$ instantaneously as soon as $f$ reaches $\pm s$. By way
of sharp contrast, at finite temperature the system undergoes
sliding even when the applied force $f(t)$ remains strictly below
$s$ in magnitude at all times. In addition, the standard deviation
$\sigma$ from the mean path is predicted to be proportional to
temperature. By way of illustration of this behavior, the
trajectories of the system corresponding to applied cyclic loads
of the form $f(t) = A \, \sin(\omega t)$ are shown in
Fig.~\ref{fig:dry} in terms of the normalized variables: $\bar{q}
= q / ( k_B T / s )$, $\bar{f} = f / s$, and $\bar{t} = \omega t$.
The ability of the theory to allow for \emph{hysteresis} and to
predict its temperature dependence is particularly noteworthy. As
expected, the response of the system is predicted to soften with
increasing temperature, in keeping with numerical simulations of
magnetic systems \cite{nowak:1997}.

\section*{Acknowledgments}

The support of the DoE through Caltech's ASCI Center for the
Simulation of the Dynamic Response of Materials is gratefully
acknowledged.



\end{document}